\documentclass[10pt,journal,twocolumn]{IEEEtran} 

\ifCLASSINFOpdf
\else
\fi

\usepackage{setspace}
\usepackage{bbm}
\usepackage[cmex10]{amsmath}
\usepackage{amssymb}
\usepackage{cite}
\usepackage{graphicx}
\usepackage{array,color}
\usepackage{amsmath}
\allowdisplaybreaks
\usepackage{stfloats}
\usepackage{graphicx}
\usepackage{epstopdf}
\usepackage{subfigure}
\usepackage{tabularx}
\usepackage{epsfig,epsf,color,balance,cite}
\usepackage{algorithmic}
\usepackage{algorithm}
\usepackage{url}
\usepackage{bm} 	  
\usepackage{multirow}
\usepackage{verbatim}  
\usepackage{cases}

\usepackage{amsthm}

\ifodd 1
\usepackage{soul}
\usepackage{color}
\setstcolor{red}

\newcommand{\mrm}[1]{{\mathrm{#1}}}
\newcommand{\bsym}[1]{{\boldsymbol{#1}}}
\newcommand{\mcal}[1]{{\mathcal{#1}}}
\newcommand{\mbb}[1]{{\mathbb{#1}}} 

\newcommand{\del}[1]{\st{#1}} 

\newcommand{\com}[1]{\textbf{\color{red} (COMMENT: #1)}} 
\newcommand{\response}[1]{\textbf{\color{green} (RESPONSE: #1)}} 
\else

\newcommand{\mrm}[1]{#1}
\newcommand{\bsym}[1]{#1}
\newcommand{\mathcal}[1]{#1}
\newcommand{\mathbb}[1]{#1}
\newcommand{\bsym}[1]{#1}

\newcommand{\del}[1]{}

\newcommand{\com}[1]{}
\newcommand{\comg}[1]{}
\newcommand{\response}[1]{}
\fi

\vspace{-0.6cm} 
	\title{\LARGE{Rotatable Antenna-Enhanced Beamforming: Signal Enhancement and Interference Suppression}} 
	\author{
			Jie~Feng, 
			Zhenbing~Liu, 
			Junjie~Dai,
			Hongbin~Chen,   
			and 
			Fangjiong~Chen
	\vspace{-1cm}
	
	\thanks{J. Feng, J. Dai, and H. Chen are with the School of Information and Communication, Guilin University of Electronic Technology, Guilin, 541004, China. (jiefeng@guet.edu.cn; jdaiguet@163.com; chbscut@guet.edu.cn)}
	\thanks{Z. Liu is with the School of Artificial Intelligence, Guilin University of Electronic Technology, Guilin, 541004, China. (zbliu@guet.edu.cn)} 
	\thanks{F. Chen is with the School of Electronic and Information Engineering, South China University of Technology, Guangzhou 510640, China. (eefjchen@scut.edu.cn). 		
	}

	} 

\begin{document}
	
	\maketitle 

 
	\begin{abstract} 
		Conventional beamforming with fixed-orientation antenna (FOA) arrays may struggle to effectively enhance signal and/or suppress interference due to significant variations in antenna directive gains over different steering angles. 
		To break this limitation, we investigate in this paper the rotatable antenna (RA)-enhanced single/multi-beam forming by exploiting the new spatial degrees of freedom (DoFs) via antennas' rotation optimization.
		Specifically, the antenna rotation angle vector (ARAV) and antenna weight vector (AWV) are jointly optimized to maximize the minimum array gain over signal directions, subject to a given constraint on the maximum array gain over interference directions.
		For the special case of single-beam forming without interference, the optimal ARAV is derived in closed-form with the maximum ratio combining (MRC) beamformer applied to the AWV. 
		For the general case of multi-beam forming, we propose an efficient alternating optimization (AO) algorithm to find a high-quality suboptimal solution by iteratively optimizing one of the ARAV and AWV with the other being fixed.
		Simulation results demonstrate that the proposed RA-based scheme can significantly outperform the traditional FOA-based and isotropic antenna (IA)-based schemes in terms of array gain. 
	\end{abstract} 
	\IEEEpeerreviewmaketitle
	
	\vspace{-0.2cm}
	\section{Introduction}
	\vspace{-0.1cm} 

	\IEEEPARstart{B}{eamforming} is an effective signal processing technique for improving the signal transmissions in multiple-antenna communication systems. 
	By manipulating the antennas' amplitudes and/or phases, the signals at different antennas can be constructively superimposed to effectively enhance the signal power and/or suppress the interference power in specific directions.
	However, conventional fixed-orientation antenna (FOA) arrays have a static antenna radiation pattern once produced, resulting in significant variations in antenna directive  gain over different steering angles \cite{2025_MA_survey}.
	In particular, when the angle of arrival/departure (AoA/AoD) of a signal (or interference) is far from (or close to) the antenna's boresight, the signal (or interference) undergoes a severe attenuation (or undesired enhancement) due to the limited spatial degree of freedom (DoF), leading to a degradation in communication performance, as can be inferred from Fig. \ref{fig_rad_gain_vs_angle}.
	
	Recently, rotatable antenna (RA), a simplified implementation of the six-dimensional movable antenna (6DMA) \cite{2025_shao_6DMA}, has emerged as a promising technology to improve communication performance.
	Through the lightweight rotation adjustments, RAs can reconfigure more favorable radiation patterns to selectively boost signal power and/or suppress interference power in specific directions \cite{2019_antenna_radiation_model}.
	It is worth pointing out that the authors in \cite{2024_Ma_multibeam_forming} proposed a movable antenna (MA)-enhanced system, which is capable of effectively increaseing/decreasing beamforming gains in specific directions.
	In contrast, owing to the remarkable antenna directive gain brought by rotation optimization, rotating antennas to improve performance is more efficient than relocating them across a wide area \cite{2025_Feng_6D_IRS}. 
	Due to its appealing advantages, RA has been studied for being integrated into various wireless systems \cite{2025_wu_RA_multi_user,2025_zheng_RA,2025_dai_RA_security, 2025_you_RA_ISAC, 2025_zheng_ISAC}.  
	However, these works adopted an ideal radiation pattern model to RAs, including the isotropic or Cosine pattern model. 
	In fact, a realistic antenna has a nonideal radiation pattern, which can be typically modeled based on the 3GPP element pattern (3GPP-EP) \cite{3GPP}. 
	The solutions of the existing works may not be directly applicable to systems with practical 3GPP-EP-based-RA arrays. 
	Besides, the performance improvement on single/multi-beam forming with an 3GPP-EP-based RA array has not been investigated, to the best of our knowledge.

	Motivated by the above, we study in this paper the RA-enhanced single/multi-beam forming by jointly optimizing the antenna rotation angle vector (ARAV) and the antenna weight vector (AWV). 
	We aim to maximize the minimum array gain (i.e., max-min array gain) over signal (desired) directions, subject to a given constraint on the maximum array gain over interference (undesired) directions. 
	The formulated optimization problem is highly non-convex with respect to (w.r.t.) both the ARAV and AWV. 
	For the special case of single-beam forming without interference, we first obtain the optimal AWV via maximum ratio combining (MRC) and then derive the optimal ARAV in closed-form. 
	For the general case of multi-beam forming, an alternating optimization (AO) algorithm is proposed to iteratively optimize one of the ARAV and AWV with the other being fixed, where the antenna rotation and beamforming optimization subproblems are suboptimally solved by using the particle swarm optimization (PSO) and the convex relaxation technique, respectively.   
	Simulation results show that the proposed RA-based scheme can significantly outperform the conventional FOA-based and isotropic antenna (IA)-based schemes in terms of array gain. 
  
	\vspace{-0.3cm}
	\section{System Model and Problem Formulation}   
	\vspace{-0.1cm} 
	
	\setlength\abovedisplayskip{3.3pt}  
	\setlength\belowdisplayskip{3.3pt} 
   
   	\begin{figure}[tp]
   	\centering 
   	\includegraphics[width=8.5cm]{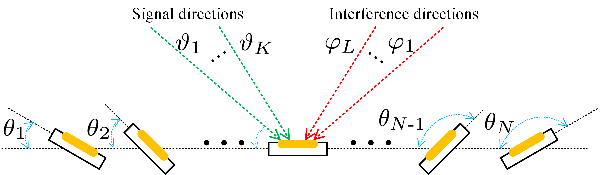} 
   	\caption{Illustration of the linear RA array.}
   	\vspace{-0.6cm}
   	\label{fig_system_model} 
   \end{figure} 
   
   \begin{figure}[tp]
	   	\centering 	   	
	   	\includegraphics[width=8cm]{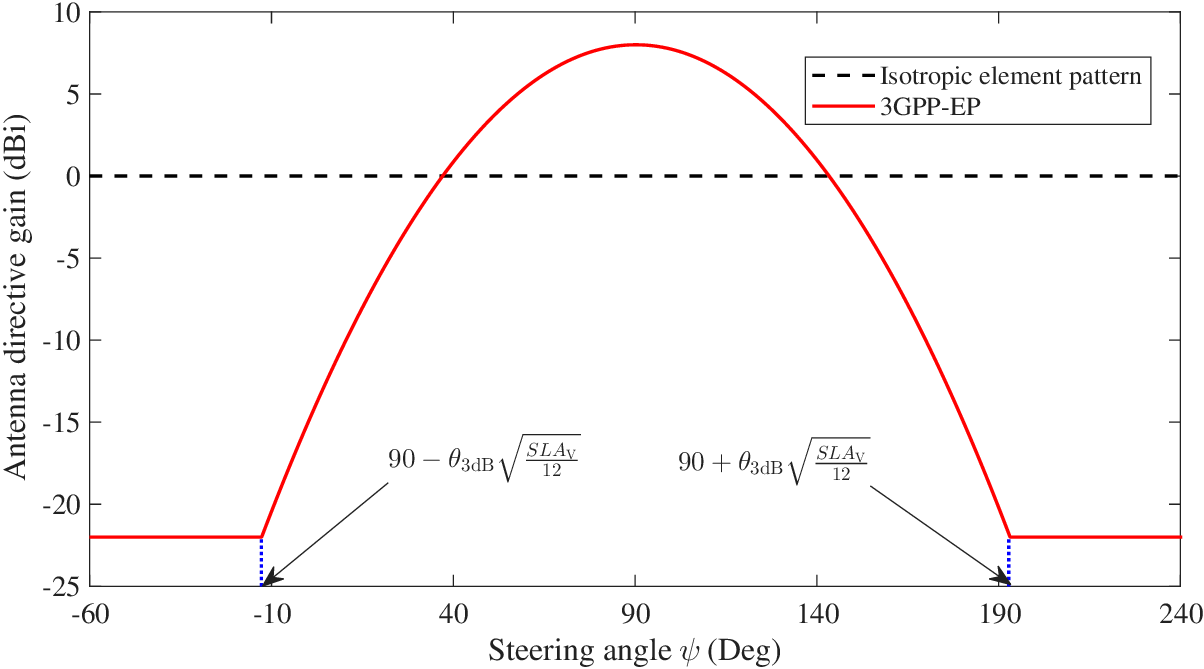} 
	   	\vspace{-0.2cm}
	   	\caption{Illustrations of the vertical antenna directive gains under different antenna radiation patterns, w.r.t. steering angle $\psi$.}
	   	\vspace{-0.4cm}
	   	\label{fig_rad_gain_vs_angle} 
   \end{figure}
  
	As shown in Fig. \ref{fig_system_model}, we consider a linear RA array with $N$ RAs, where each RA with 3GPP-EP can be flexibly rotated to selectively enhance the array gain over desired signal directions and/or suppress the array gain in undesired interference directions.

	In this paper, we consider a practical 3GPP-EP for each RA \cite{3GPP}, defined as 
	\begin{align} \label{eqn_ele_pat_gain}
		G_{\mrm{e}} (\psi) = G_{\mrm{e}}^{\max} - \min\left\{ -G_{\mrm{E,V}}(\psi), A_{\max}  \right\} ,
	\end{align} 
	where  
	$G_{\mrm{e}}^{\max} = 8 $ dBi denotes the maximum directional gain of each RA,
	$G_{\mrm{E,V}}(\psi)$ denotes the vertical radiation pattern,
	and $A_{\max}= 30$ dB denotes the front-to-back ratio.\footnote{In this paper, we only consider the vertical radiation pattern due to space limitation, and this case can be extended to the more general case of the combination of the horizontal and vertical radiation patterns in the future work.}
	In the above, the vertical radiation pattern in dBi is computed by 
	\begin{align} \label{eqn_hor_rad_pat}
		G_{\mrm{E,V}}(\psi) = -\min \left\lbrace 12 \left(\frac{ \psi - 90^{\circ}}{\theta_{\mrm{3dB}}}\right)^2, SLA_{\mrm{V}} \right\rbrace , 
	\end{align}
	where $\theta_{\mrm{3dB}} = 65^{\circ}$ denotes the vertical $3$ dB beamwidth 
	and $SLA_{\mrm{V}} = 30$ dB denotes the side-lobe level limit.
	To illustrate the impact brought by different incoming angles on the antenna directive gain, in Fig. \ref{fig_rad_gain_vs_angle}, we plot the vertical antenna directive gain versus steering angles.
	It is obseved from Fig. \ref{fig_rad_gain_vs_angle}, the antenna pattern gain varies significantly within the range of steering angles $\left[\left(90- \theta_{\mrm{3dB}}\sqrt{\frac{SLA_{\mrm{V}}}{12}} \right)^{\circ}, \left( 90 + \theta_{\mrm{3dB}}\sqrt{\frac{SLA_{\mrm{V}}}{12}} \right)^{\circ}\right] $, which demonstrates that the necessity of the antenna rotation optimization.
	Based on the above, to achieve an efficient rotation optimization, we add a constraint to each RA, which is expressed as 
	\begin{align}
		\theta_{n} \in \left[ \theta_{\min} , \theta_{\max} \right], n\in \mcal{N},
	\end{align}
	where $\theta_{\min} \triangleq \left(90-\theta_{\mrm{3dB}}\sqrt{\frac{SLA_{\mrm{V}}}{12}} \right)^{\circ}$ 
	and $\theta_{\max} \triangleq \left( 90 + \theta_{\mrm{3dB}}\sqrt{\frac{SLA_{\mrm{V}}}{12}} \right)^{\circ}$.
	
	Let $\bsym{\theta} = \left[ \theta_{1},...,\theta_{N} \right]^{\top}  \in \mbb{R}^{N \times 1}$ denote the ARAV, 	
	where $\theta_{n}$ denotes the $n$-th RA's rotation angle, $n \in \{1,...,N\} \triangleq \mcal{N}$.
	Based on the above, the effective antenna directive gain vector in the linear scale w.r.t. a steering angle $\psi$ is given by 
	\begin{align}
		\bsym{g} \left( \bsym{\theta}, \psi \right) = \left[ \sqrt{\bar{G}_{\mrm{e}}(\psi - \theta_{1})}, ..., 
		\sqrt{\bar{G}_{\mrm{e}}(\psi - \theta_{N})} \right]^{\top} .
	\end{align}
	where $\bar{G}_{\mrm{e}}(\phi) = 10^{\frac{G_{\mrm{e}}(\phi)}{10}}$ denotes the effective antenna directive gain in the linear scale.
	 
	The steering vector of the RA array w.r.t. the AoA $\psi \in [0^{\circ},180^{\circ}]$ is expressed as 
	\begin{align}
		\bsym{a}(\psi) = & \left[  1, e^{\jmath \frac{2\pi d}{\lambda} \cos(\psi) }, ...,  e^{\jmath \frac{2\pi d}{\lambda} (N-1) \cos(\psi)  } \right]^{\top} ,
	\end{align} 
	where 
	$d$ denotes the distance between two adjacent antennas
	and $\lambda$ denotes the signal wavelength.  
	
	With the above setups, the array gain over $\psi$ is given by 
	\begin{align}
		G_{\mrm{Arr}}(\bsym{w},\bsym{\theta},\psi) & = 
		\left| \bsym{w}^{H} \left[ \bsym{g}(\bsym{\theta}, \psi) \odot \bsym{a}(\psi) \right] \right|^2   \\
		& = \left| \bsym{g}^{\top}(\bsym{\theta}, \psi) \text{diag} \left( \bsym{w}^{H} \right) \bsym{a}(\psi) \right|^2,
	\end{align}
	where $\bsym{w} \in \mbb{C}^{N \times 1}$ denotes the AWV for beamforming with $\|\bsym{w}\|_{2} \leq 1$,
	and $\odot$ denotes the Hadamard product.
	
	As shown in Fig. \ref{fig_system_model}, we consider the interference dominant scenario (thus, the receiver noise can be neglected) in which the signals (over desired directions $\{\vartheta_{k}\}_{k=1}^{K}$) and the interferences (over undesired directions $\{\varphi_{l}\}_{l=1}^{L}$) coexist.
	Under this setup, our objective is to maximize the minimum array gain over $\{\vartheta_{k}\}_{k=1}^{K}$ (denoted by $G_{\min}$) while minimizing the maximum array gain over $\{\varphi_{l}\}_{l=1}^{L}$ (denoted by $\eta_{\max}$), by jointly optimizing the AWV $\bsym{w}$ and the ARAV $\bsym{\theta}$.
	Accordingly, the optimization problem can be formulated as 
	\begin{align}
		\text{(P1)}: \max_{\bsym{w}, \bsym{\theta}} ~& G_{\min} \label{P1} \\
				\text{s.t.}	~ & G_{\mrm{Arr}}(\bsym{w},\bsym{\theta},\vartheta_{k}) \geq G_{\min}, ~ k \in \mcal{K}, \tag{\ref{P1}a} \label{P1_a} \\
				& G_{\mrm{Arr}}(\bsym{w},\bsym{\theta},\varphi_{l}) \leq \eta_{\max}, ~ l \in \mcal{L}, \tag{\ref{P1}b} \label{P1_b} \\
				& \theta_{n} \in [\theta_{\min} , \theta_{\max}], ~ n \in \mcal{N},\tag{\ref{P1}c} \label{P1_c}  \\
				& \|\bsym{w}\|_{2} \leq 1,  \tag{\ref{P1}d} \label{P1_d}  
	\end{align}
	where constraint \eqref{P1_c} ensures that the RAs are efficiently rotated within the given spatial range, and constraint \eqref{P1_d} ensures that the AWV power is normalized to be no larger than one.
	
	\vspace{-0.2cm}	
	\section{Solution}
	\vspace{-0.1cm}
	
	\subsection{The Special Case of Single-Beam Forming} 
	 
	 In this section, we consider the single-beam forming and no interference setups, i.e., $K = 1$ and $L = 0$, to draw important insights into (P1).  
	 In this case, no interference is present, and thus (P1) is simplified to (by dropping the signal index)
	 \begin{align}
	 	\text{(P2)}: \max_{\bsym{w}, \bsym{\theta}} ~& G_{\mrm{Arr}}(\bsym{w},\bsym{\theta},\vartheta) \label{P2} \\
	 	\text{s.t.}	~ & \theta_{n} \in [\theta_{\min} , \theta_{\max}], ~ n \in \mcal{N},\tag{\ref{P2}a} \label{P2_a} \\ 
	 	& \|\bsym{w}\|_{2} \leq 1. \tag{\ref{P2}b} \label{P2_b}  
	 \end{align}
	 For any given ARAV $\bsym{\theta}$, it is known that the maximum ratio combining (MRC) beamformer is the optimal receive beamforming solution to problem (P2), which is given by 
	 \begin{align} \label{eqn_opt_beamforming}
	 	\bsym{w}^{\star} = \frac{\bsym{a}(\vartheta)}{\|\bsym{a}(\vartheta)\|_{2}}. 
	 \end{align}
	 Then, by substituting $\bsym{w}^{\star}$ into the objection function of problem (P2), we have
	 \begin{align}
	 	 G_{\mrm{Arr}}(\bsym{w}^{\star},\bsym{\theta},\vartheta)
	 	 = & \left| \bsym{g}^{\top}(\bsym{\theta}, \vartheta) \text{diag} \left( (\bsym{w}^{\star})^{H} \right) \bsym{a}(\vartheta) \right|^2 \\
	 	 = & \frac{1}{N} \left| \bsym{g}^{\top}(\bsym{\theta}, \vartheta)  \bsym{1}_{N} \right|^2 \triangleq G_{\mrm{Arr}}(\bsym{\theta},\vartheta),  
	 \end{align}
	 where $\bsym{1}_{N}$ is an $N$-dimensional all-ones vector.
	Based on the above, problem (P2) can be simplified as the maximization of the directive gains of all the RAs, which is expressed as
	\begin{align}
		\text{(P2.1)}: \max_{ \bsym{\theta}} ~& G_{\mrm{Arr}}(\bsym{\theta},\vartheta) \label{P2_1} \\
		\text{s.t.}	~ & \theta_{n} \in [\theta_{\min} , \theta_{\max}], ~ n \in \mcal{N}.\tag{\ref{P2_1}a} \label{P2_1_a} 
	\end{align}
	 To achieve \eqref{P2_1}, the directive gain of each RA should be maximized, which requires that the boresight direction of each RA is aligned with the signal direction.
	 As such, the optimal solution to problem (P2.1) is given by 
	 \begin{align} \label{eqn_opt_angle}
	 	\theta_{n}^{\star} = \vartheta - 90^{\circ},  n\in\mcal{N}.
	 \end{align}  
	 By substituting \eqref{eqn_opt_angle} into \eqref{P2_1} yields the maximum array gain for single-beam forming, given by  
	 \begin{align} \label{eqn_full_arr_gain}
	 	G_{\mrm{Arr}}(\{\theta_{n}^{\star}\},\vartheta) = \overbrace{\underbrace{N}_{\text{Full beamforming gain}} 
	 	\times \underbrace{10^{\frac{G_{\mrm{e}}^{\max}}{10}}}_{\text{Full antenna directive gain}}}^{\text{Full array gain}}.
	 \end{align}
	 
	\vspace{-0.2cm}
	\subsection{The General Case of Multi-Beam Forming}
	\vspace{-0.1cm}
	
	In this subsection, we consider the general case of multi-beam forming and propose an AO algorithm to alternately optimize the AWV and the ARAV of the RA array in an iterative manner, whose details are as follows.

	\subsubsection{Optimization of $\bsym{w}$ With Given $\bsym{\theta}$}
	For any given ARAV $\bsym{\theta}$, problem (P1) is reduced to 
	\begin{align}
		\text{(P3)}: \max_{\bsym{w}} ~& G_{\min} \label{P3} \\ 
		\text{s.t.}	~ & G_{\mrm{Arr}}(\bsym{w},\bsym{\theta},\vartheta_{k}) \geq G_{\min}, ~ k \in \mcal{K}, \tag{\ref{P3}a} \label{P3_a} \\
		& G_{\mrm{Arr}}(\bsym{w},\bsym{\theta},\varphi_{l}) \leq \eta_{\max}, ~ l \in \mcal{L}, \tag{\ref{P3}b} \label{P3_b} \\
		& \|\bsym{w}\|_{2} \leq 1. \tag{\ref{P3}c} \label{P3_c}   
	\end{align}
	As can be observed, since the left-hand side of \eqref{P3_a} is convex w.r.t. $\bsym{w}$, the non-convexity of \eqref{P3_a} can be tackled by applying the SCA technique, which solves problem (P3) in an iterative manner.  
	Specifically, for $\bsym{w}^{(i)} \in \mbb{C}^{N \times 1}$ obtained in the $i$-th iteration of SCA,  we can construct the following surrogate function $\bar{G}_{\mrm{Arr}}(\bsym{w},\bsym{\theta},\vartheta_{k}|\bsym{w}^{(i)})$ to globally minorize $G_{\mrm{Arr}}(\bsym{w},\bsym{\theta},\vartheta_{k})$ by applying the first-order Taylor expansion at $\bsym{w}^{(i)}$, i.e.,
	\begin{align}
		& G_{\mrm{Arr}}(\bsym{w},\bsym{\theta},\vartheta_{k}) 
		\geq  \bar{G}_{\mrm{Arr}}(\bsym{w},\bsym{\theta},\vartheta_{k}|\bsym{w}^{(i)}) 	\nonumber \\
		= &  2 \Re \left\{ \left( \bsym{w}^{(i)} \right)^{H} \bsym{v}(\bsym{\theta}, \vartheta) \bsym{v}^{H}(\bsym{\theta}, \vartheta) \left(  \bsym{w} - \bsym{w}^{(i)} \right)  \right\}  \nonumber \\
		&  + G_{\mrm{Arr}}(\bsym{w}^{(i)},\bsym{\theta},\vartheta_{k})   \nonumber \\
		\triangleq & 2 \Re \left\{ \left( \bsym{w}^{(i)} \right)^{H} \bsym{v}(\bsym{\theta}, \vartheta) \bsym{v}^{H}(\bsym{\theta}, \vartheta) \bsym{w} \right\}  \nonumber \\   
		&  - G_{\mrm{Arr}}(\bsym{w}^{(i)},\bsym{\theta},\vartheta_{k}) . \label{eqn_reformulated_gain}
	\end{align} 
	where $\bsym{v}(\bsym{\theta}, \vartheta) \triangleq \bsym{g}(\bsym{\theta}, \vartheta) \odot \bsym{a}(\bsym{\theta}, \vartheta)$,
	and $\Re\{ \cdot \}$ denotes the real part of a complex number.
	Then, replacing $G_{\mrm{Arr}}(\bsym{w},\bsym{\theta},\vartheta_{k})$ in constraint \eqref{P3_a} with the surrogate function $\bar{G}_{\mrm{Arr}}(\bsym{w},\bsym{\theta},\vartheta_{k}|\bsym{w}^{(i)})$ in \eqref{eqn_reformulated_gain} yields the following linear constraint:
	\begin{align}
		\bar{G}_{\mrm{Arr}}(\bsym{w},\bsym{\theta},\vartheta_{k}|\bsym{w}^{(i)}) \geq G_{\min}, ~ k \in \mcal{K}.   \label{eqn_discrete_gain_AA_linear}
	\end{align}  
	As a result, for a given $\bsym{w}^{(i)}$ obtained in the $i$-th iteration, the corresponding problem in the ($i$+$1$)-th iteration can be formulated as 
	\begin{align}
	\text{(P3.1)}: \max_{\bsym{w}} ~& G_{\min} \label{P3_1} \\
	\text{s.t.}	~ & \bar{G}_{\mrm{Arr}}(\bsym{w},\bsym{\theta},\vartheta_{k}) \geq G_{\min}, ~ k \in \mcal{K}, \tag{\ref{P3_1}a} \label{P3_1_a} \\
	& G_{\mrm{Arr}}(\bsym{w},\bsym{\theta},\varphi_{l}) \leq \eta_{\max}, ~ l \in \mcal{L}, \tag{\ref{P3_1}b} \label{P3_1_b} \\
	& \|\bsym{w}\|_{2} \leq 1. \tag{\ref{P3_1}c} \label{P3_1_c}   
	\end{align}	
	Since constraint \eqref{P3_1_a} is linear and constraints \eqref{P3_1_b} and \eqref{P3_1_c} are convex quadratic w.r.t. $\bsym{w}$, problem (P3.1) is convex and thus can be solved via CVX \cite{grant2014cvx}.
	The detailed steps of the proposed algorithm for solving problem (P3) is summarized in \textbf{Algorithm \ref{Algorithm_1}}.  

 	\begin{algorithm}[t]  
		\caption{Proposed Algorithm for Solving (P3)}  \label{Algorithm_1}
		\begin{algorithmic}[1] 
			\STATE \textbf{Input}: $\bsym{w}^{(0)}$, $\bsym{\theta}$, $\{\vartheta_{k}\}$, $\{\varphi_{l}\}$, $\eta_{\max}$, and threshold $\delta_{\mrm{th}}$.
			
			\STATE $i = 0$.  
			
			\REPEAT  
			
			\STATE Update $i = i + 1$. 
			
			\STATE Obtain $\bsym{w}^{(i)}$ by solving problem (P3.1) with $\bsym{w}^{(i-1)}$.  
				
			\UNTIL Increase of \eqref{P3_1} is lower than $\delta_{\mrm{th}}$ 
			
			\RETURN $\bsym{w}^{(i)}$.
			
		\end{algorithmic}
	\end{algorithm}

	\subsubsection{Optimization of $\bsym{\theta}$ With Given $\bsym{w}$}

	For any given AWV $\bsym{w}$, problem (P1) is reformulated as
	\begin{align}
		\text{(P4)}: \max_{\bsym{\theta}} ~& G_{\min} \label{P4} \\
		\text{s.t.}	~ & G_{\mrm{Arr}}(\bsym{w},\bsym{\theta},\vartheta_{k}) \geq G_{\min}, ~ k \in \mcal{K}, \tag{\ref{P4}a} \label{P4_a} \\
		& G_{\mrm{Arr}}(\bsym{w},\bsym{\theta},\varphi_{l}) \leq \eta_{\max}, ~ l \in \mcal{L}, \tag{\ref{P4}b} \label{P4_b} \\
		& \theta_{n} \in [\theta_{\min} , \theta_{\max}], ~ n \in \mcal{N}. \tag{\ref{P4}c} \label{P4_c}  
	\end{align}
	As can be observed, problem (P4) is difficult to be solved via standard convex optimization methods due to its high non-convexity.
	To efficiently address this problem, we adopt the particle swarm optimization (PSO) algorithm \cite{2004_PSO}, whose details are as follows.
	
	The PSO algorithm starts by randomly generating $S$ particles, each of which is expressed as
	\begin{align}
		\bsym{\theta}_{s}^{(0)} = \left[ \theta_{1}^{(0)}, ..., \theta_{N}^{(0)} \right], 
	\end{align} 
	with $s \in \{1,...,S\} \triangleq \mcal{S}$, where $\theta_{n}^{(0)} \in \left[\theta_{\min}, \theta_{\max}\right] $, $n \in \mcal{N}$.
	The $s$-th particle updates its position according to the individual experience (i.e., the known local best position, denoted by $\bsym{\theta}_{s,\mrm{Loc}}^{\star}$) and the swarm experience (i.e., the known global best position, denoted by $\bsym{\theta}_{\mrm{Glo}}^{\star}$).
	At the $t$-th iteration, the velocity and the position of the $s$-th particle are respectively calculated by 
	\begin{align}
		\bsym{v}_{s}^{(t)} = & w^{(t)} \bsym{v}_{s}^{(t)} + s_{\mrm{Loc}} \mrm{rand}() \left( \bsym{\theta}_{s,\mrm{Loc}}^{\star} - \bsym{\theta}_{s}^{(t-1)}\right) \nonumber \\ 
							& + s_{\mrm{Glo}} \mrm{rand}() \left( \bsym{\theta}_{\mrm{Glo}}^{\star} - \bsym{\theta}_{s}^{(t-1)}\right), \label{eqn_velocity} \\
		\bsym{\theta}_{s}^{(t)} = & f_{\text{Pro}} \left( \bsym{\theta}_{s}^{(t-1)} + \bsym{v}_{s}^{(t)} \right), \label{eqn_position} 			
	\end{align}
	$s \in \mcal{S}$, 
	where $w$ denotes the inertia weight, 
	$s_{\mrm{Loc}}$ and $s_{\mrm{Glo}}$ denote the individual and global learning factors, respectively,
	$\mrm{rand}()$ denotes the random number function which randomly returns a number between zero and one with uniform distribution,
	and $f_{\text{Pro}}(\cdot)$ denotes the project function which is to guarantee constraint \eqref{P4_c}. 
	In the above, the inertia weight at the $t$-th iteration is computed by 
	\begin{align} \label{eqn_inertia_weight}
		w^{(t)} = w_{\mrm{Ini}} - \frac{(w_{\mrm{Ini}} - w_{\mrm{Final}})t}{T_{\max}}.
	\end{align}
	where $T_{\max}$ denotes the maximum iteration number, 
	and $w_{\mrm{Ini}}$ and $w_{\mrm{Final}}$ denote the initial and final inertia weights, respectively.
	The project function $f_{\text{Pro}}( \bsym{\theta} )$ is given by 
	\begin{align}
		\left[ f_{\text{Pro}}( \bsym{\theta} ) \right]_{n} = 
		\begin{cases}
			\theta_{\min}, \quad \text{if} \left[\bsym{\theta}\right]_{n} <  \theta_{\min},  \\
			\theta_{\max}, \quad \text{if} \left[\bsym{\theta}\right]_{n} >  \theta_{\max},  \\
			\left[\bsym{\theta}\right]_{n}, \quad \text{otherwise}, \\
		\end{cases} n \in \mcal{N},
	\end{align}
	where $\left[\bsym{\theta}\right]_{n}$ denotes the $n$-th entry of vector $\bsym{\theta}$.
	Besides, at the $t$-th iteration, the fitness function of the $s$-th particle is calculated by 
	\begin{align} \label{eqn_fitness_function}
		f_{\text{Fit}} \left( \bsym{\theta}_{s}^{(t)} \right) = G_{s,\min}^{(t)}  - \rho.
	\end{align}
	where $G_{s,\min}^{(t)} = \min\left( \left\{G_{\mrm{Arr}}(\bsym{w},\bsym{\theta}_{s}^{(t)},\vartheta_{k})| k \in \mcal{K} \right\} \right) $ denotes the minimum array gain over signal directions w.r.t. $\bsym{\theta}_{s}^{(t)}$,
	and $\rho$ denotes an adaptive penalty factor to guarantee constraint \eqref{P4_b} which is computed as	
	\begin{align}
		\rho = \tau \sum_{\stackrel{l \in \mcal{L}}{G_{\mrm{Arr}}(\bsym{w},\bsym{\theta}_{s}^{(t)},\varphi_{l}) > \eta_{\max}}} \hspace{-1cm} G_{\mrm{Arr}}(\bsym{w},\bsym{\theta}_{s}^{(t)},\varphi_{l}),
	\end{align}
	with $\tau$ denoting a large positive penalty parameter.
	The details of the PSO-based algorithm is provided in \textbf{Algorithm \ref{Algorithm_2}}.

 	\begin{algorithm}[t]  
		\caption{Proposed Algorithm for Solving (P4)}  \label{Algorithm_2}
		\begin{algorithmic}[1] 
			\STATE \textbf{Input}: $\bsym{w}$,  $\{\vartheta_{k}\}$,  $\{\varphi_{l}\}$, $\eta_{\max}$, $w_{\mrm{Ini}}$, $w_{\mrm{Final}}$,  $T_{\max}$, $F_{\mrm{Last}}^{\star}$, and $\delta_{\mrm{th}}$. 
			
			\STATE Initialize $S$ particles with positions $\bsym{\theta}^{(0)}$ and velocities $\bsym{v}^{(0)}$.
			
			\STATE Calculate the fitness values of all the particles $\left\{f_{\text{Fit}} \left( \bsym{\theta}_{1}^{(0)} \right),...,f_{\text{Fit}} \left( \bsym{\theta}_{S}^{(0)} \right)\right\}$ via \eqref{eqn_fitness_function}.
			
			\STATE Obtain $\bsym{\theta}_{\mrm{Glo}}^{\star} = \arg\max_{\bsym{\theta}_{s}^{(0)}} \left\{f_{\text{Fit}} \left( \bsym{\theta}_{1}^{(0)} \right),...,f_{\text{Fit}} \left( \bsym{\theta}_{S}^{(0)} \right)\right\}$ and $\bsym{\theta}_{s,\mrm{Loc}}^{\star} = \bsym{\theta}_{s}^{(0)}$, $s \in \mcal{S}$.  
			
			\FOR{$t=1$ to $T_{\max}$}
			
				\STATE Update $w^{(t)}$ via \eqref{eqn_inertia_weight}.
				 
				\FOR{$s=1$ to $S$}
				
					\STATE Update $\bsym{v}_{s}^{(t)}$ and $\bsym{\theta}_{s}^{(t)}$ via \eqref{eqn_velocity} and \eqref{eqn_position}, respectively.
					 
					\STATE Compute $f_{\text{Fit}} \left( \bsym{\theta}_{s}^{(t)} \right)$ via \eqref{eqn_fitness_function}.
					
					\IF{$f_{\text{Fit}} \left( \bsym{\theta}_{s}^{(t)} \right) > f_{\text{Fit}} \left( \bsym{\theta}_{s,\mrm{Loc}}^{\star} \right)$}
						
						\STATE Update $\bsym{\theta}_{s,\mrm{Loc}}^{\star} = \bsym{\theta}_{s}^{(t)} $.
					
					\ENDIF
					
					\IF{$f_{\text{Fit}} \left( \bsym{\theta}_{s}^{(t)} \right) > f_{\text{Fit}} \left( \bsym{\theta}_{\mrm{Glo}}^{\star} \right)$}
					
						\STATE Update $\bsym{\theta}_{\mrm{Glo}}^{\star} = \bsym{\theta}_{s}^{(t)} $.
						
						\STATE Update $F_{\mrm{Cur}}^{\star} = f_{\text{Fit}} \left( \bsym{\theta}_{s}^{(t)} \right) $.
					
					\ENDIF
				
				\ENDFOR
				
			 	\IF{$ F_{\mrm{Cur}}^{\star}- F_{\mrm{Last}}^{\star} < \delta_{\mrm{th}}$}
			 	
			 		\STATE $\text{\textbf{break}}$.
			 		
			 	\ELSE
			 	
			 		\STATE $F_{\mrm{Last}}^{\star} = F_{\mrm{Cur}}^{\star}$.
			 	
			 	\ENDIF
			 	
			\ENDFOR 
			  
			\RETURN $\bsym{\theta}_{\mrm{Glo}}^{\star}$.
			
		\end{algorithmic}
	\end{algorithm}
	
	\subsubsection{Overall Algorithm}
	
	The detailed steps of the proposed AO algorithm is shown in \textbf{Algorithm \ref{Algorithm_3}}.
	Since the optimal objective value of the problem (P1) is non-decreasing over iterations and is upper-bounded, the convergence of the proposed AO algorithm is guaranteed.
	The computational complexity of \textbf{Algorithm 1} and that of \textbf{Algorithm 2} are $\mcal{O}\left( T_{1} K L N^{3.5} \log\left( \epsilon \right) \right)$ and $\mcal{O}\left( T_{2} S N \right)$, respectively, where $\epsilon$ is a given solution accuracy, and $T_{1}$ and $T_{2}$ respectively denote the numbers of iterations required for convergence of \textbf{Algorithm 1} and \textbf{Algorithm 2}.
	As a result, the computational complexity of the proposed AO algorithm is $\mcal{O}\left( T_{3}   \left( T_{1} K L N^{3.5} \log\left( \epsilon \right) + T_{2} S N \right) \right)$, where $T_{3}$ denotes the number of iterations required for convergence.
	
	 	\begin{algorithm}[t]  
		\caption{Proposed AO Algorithm for Solving (P3)}  \label{Algorithm_3}
		\begin{algorithmic}[1] 
			\STATE \textbf{Input}: $\bsym{w}^{(0)}$, $\bsym{\theta}^{(0)}$, $\{\vartheta_{k}\}$, $\{\varphi_{l}\}$, $\eta_{\max}$, $w_{\mrm{Ini}}$, $w_{\mrm{Final}}$, $T_{\max}$, $F_{\mrm{Last}}^{\star}$, and $\delta_{\mrm{th}}$.
			
			\STATE $m = 0$.  
			
			\REPEAT   
			
				\STATE Update $m = m + 1$. 
			
				\STATE Obtain $\bsym{w}^{(m)}$ via \textbf{Algorithm 1} with $\bsym{\theta}^{(m-1)}$ and $\bsym{w}^{(m-1)}$.  
				
				\STATE Obtain $\bsym{\theta}^{(m)}$ via \textbf{Algorithm 2} with $\bsym{w}^{(m)}$.
					
				\UNTIL Increase of \eqref{P1} is lower than $\delta_{\mrm{th}}$ 
			
			\RETURN $\bsym{w} = \bsym{w}^{(m)}$ and $\bsym{\theta} = \bsym{\theta}^{(m)}$.
			
		\end{algorithmic}
	\end{algorithm}

	\vspace{-0.2cm}
	\section{Simulation Results}
	\vspace{-0.1cm}
	
	In this section, we present simulation results to examine the performance of the proposed RA-based scheme. 
	Unless otherwise specified, system parameters are set as follows:  
	$\bsym{\theta}^{0} = \bsym{0}_{N}$,  
	$d = \frac{\lambda}{2}$ m,  
	$N  =15 $, 
	$\eta_{\max} = -10$ dB, 
	$w_{\mrm{Ini}}=0.9$,
	$w_{\mrm{Final}}=0.2$,
	$T_{\max} = 10^{2}$, 
	$\tau= 10^{6}$,
	$F_{\mrm{Last}}^{\star} = -\inf$
	and $\delta_{\mrm{th}}=10^{-2}$.   
	Moreover, the phase shifts in the initial AWV $\bsym{w}^{0}$ are randomly generated within the range of $[0, 2\pi)$.
	For performance comparison, we consider the two benchmark schemes: 
	1) \textbf{FOA-based} scheme in which the ARAV is fixed as $\bsym{\theta} = \bsym{0}_{N}$,
	and 2) \textbf{IA-based} scheme in which the antenna directive gain is set to one. 
	To ensure statistical average, all the simulation results are averaged over $360$ times.

\begin{figure}[tp]
		\centering 
		\includegraphics[width=7.5cm]{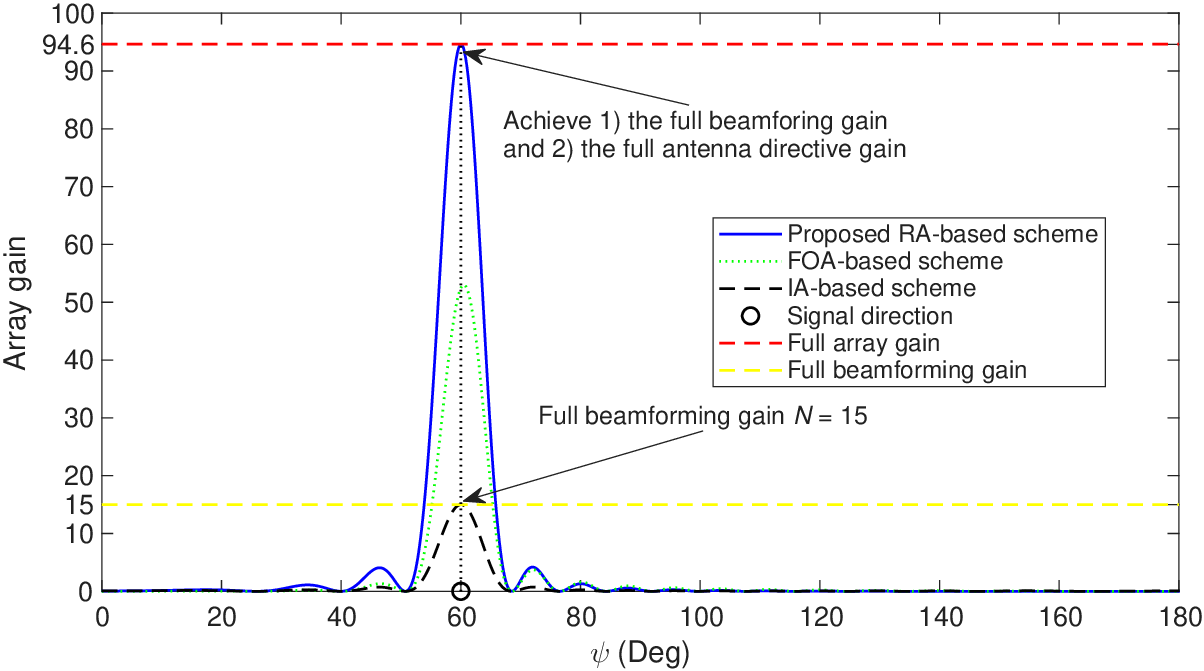} 
		\caption{Array gain versus AoA $\psi$, with $\vartheta = 60^{\circ}$.}
		   	\vspace{-0.4cm}
		\label{fig_gain_vs_angle_1} 
	\end{figure}

	Fig. \ref{fig_gain_vs_angle_1} plots the array gain versus AoA $\psi$ under different schemes. 
	It is observed that the proposed RA-based scheme significantly outperforms the benchmark schemes in terms of array gain.
	This is expected since by exploiting the additional spatial DoFs in antenna rotation optimization, the proposed scheme can enjoy a higher antenna directive gain.
	Besides, the proposed scheme achieves 1) the full beamforming gain and 2) the full antenna directive gain, i.e., the full array gain $N \times 10^{\frac{8}{10}} \approx 94.6$, as can be inferred from \eqref{eqn_full_arr_gain}.
	Moreover, the array gain of the FOA-based scheme is higher than that of the IA-based scheme.
	This is because with the 3GPP-EP, the FOA-based scheme can additionally enjoy a higher antenna directive gain than the IA-based scheme.
 	
\begin{figure}[tp]
		\centering
		\subfigure[Array gain versus AoA $\psi$, with $\vartheta_{1} = 55^{\circ}$, $\vartheta_{2} = 60^{\circ}$, $\varphi_{1} = 20^{\circ}$, and $\varphi_{2} = 160^{\circ}$.]{
			\includegraphics[width=7.5cm]{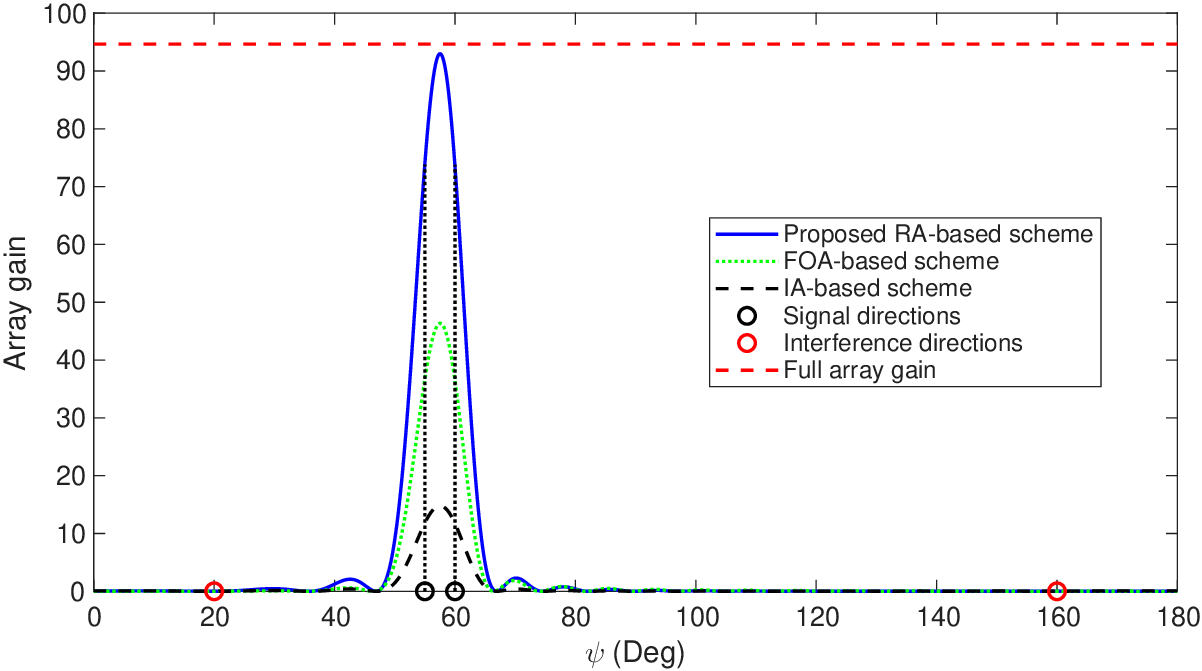} 
			\label{fig_gain_vs_angle_2a}  }   
		\subfigure[Array gain versus AoA $\psi$, with $\vartheta_{1} = 60^{\circ}$, $\vartheta_{2} = 140^{\circ}$, $\varphi_{1} = 20^{\circ}$, and $\varphi_{2} = 160^{\circ}$.]{
			\includegraphics[width=7.5cm]{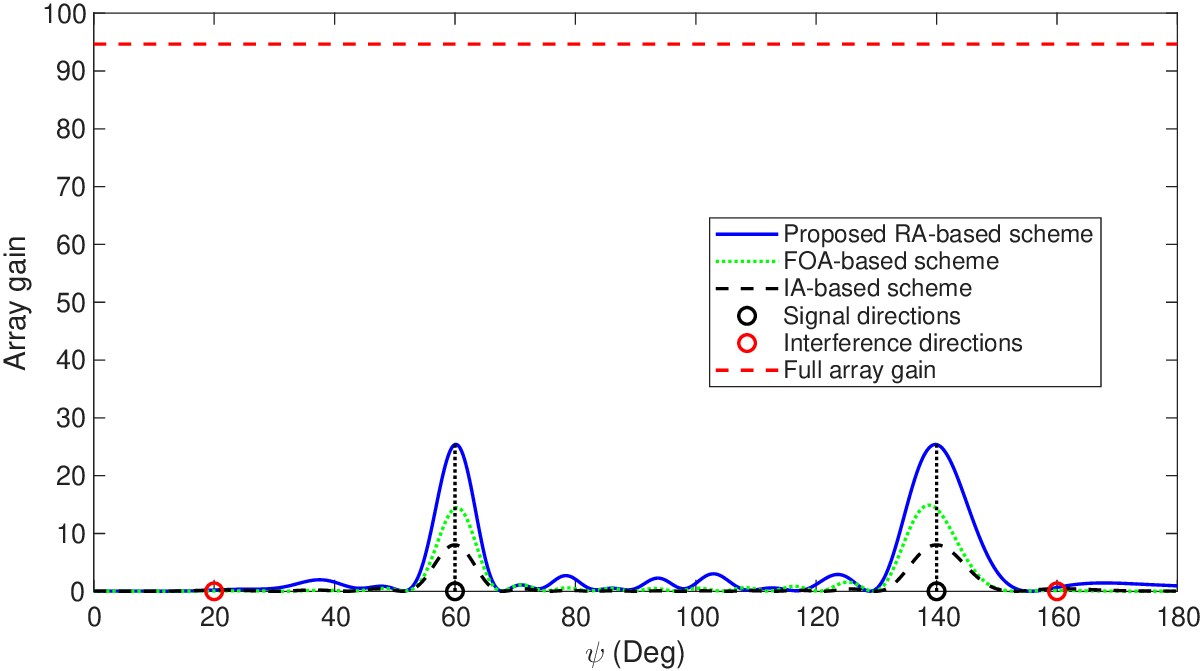} 
			\label{fig_gain_vs_angle_2b}  } 
		\caption[]{Comparison of array gain patterns under different setups.}
		\vspace{-0.4cm}
		\label{fig_simulations} 
	\end{figure}

	Fig. \ref{fig_simulations} shows the array gain patterns under different schemes and simulation parameters.
	Some interesting observations are made as follows.
	First, the proposed RA-based scheme achieves a much higher array gain over the signal (desired) directions than the benchmark schemes.
	Second, all the schemes have effective interference suppression over the interference (undesired) directions, i.e., below the predefined threshold $\eta_{\max}$, which validates the effectiveness of the proposed AO algorithm. 
	Finally, the proposed scheme in Fig. \ref{fig_gain_vs_angle_2a} can achieve $78.16\%$ of the full array gain, while that in Fig. \ref{fig_gain_vs_angle_2b} achieves $26.88\%$ of the full array gain. 
	This is since the former can share a higher antenna directive gain over the adjacent signal directions than the latter due to the closer signal directions, which demonstrates the importance of antenna rotation optimization.

\begin{figure}[tp]
		\centering 
		\includegraphics[width=7.5cm]{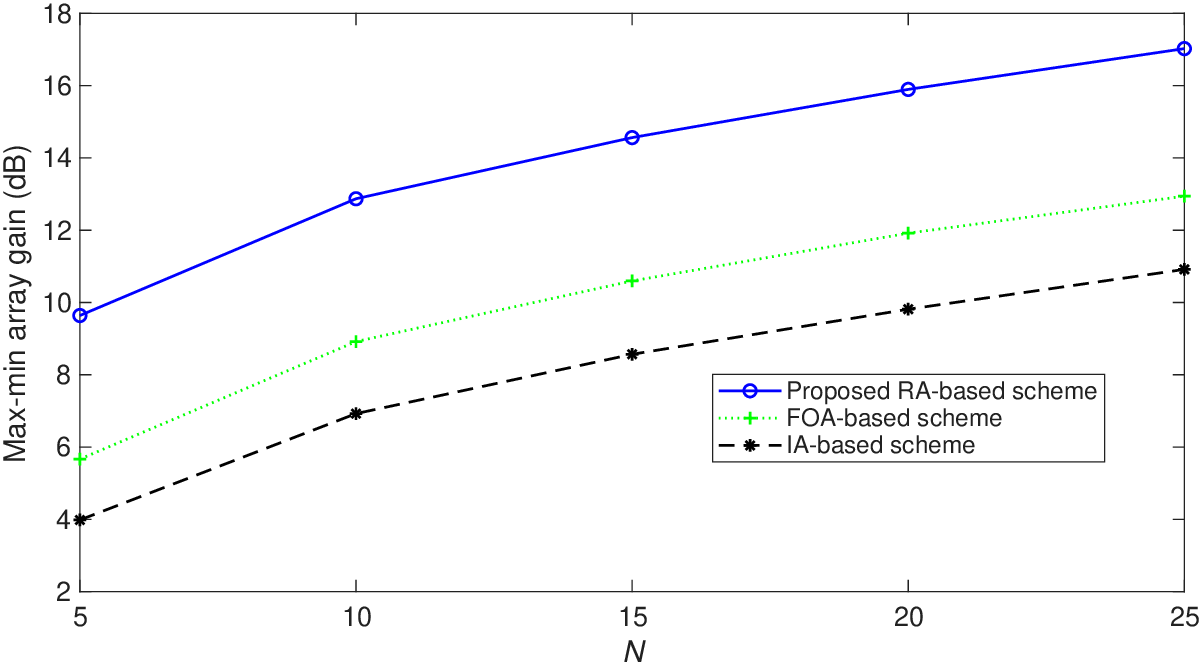} 
		\caption{Max-min array gain versus number of RAs, $N$.}
	   	\vspace{-0.6cm}
		\label{fig_gain_vs_antenna} 
	\end{figure}

	Fig. \ref{fig_gain_vs_antenna} depicts the max-min array gain versus the number of RAs, $N$.
	The signal directions $\{\vartheta_{k}\}_{k=1}^{K}$ and the interference directions $\{\varphi_{l}\}_{l=1}^{L}$ are randomly generated within $[0, 180^{\circ}]$ with uniform distribution.
	It is observed that the max-min array gains of all the schemes significantly increase with $N$.
	Besides, the proposed scheme significantly outperforms the benchmark schemes in terms of the max-min array gain.
	Specifically, the proposed scheme has about $6$ dB and $4$ dB performance improvement over the FOA-based scheme and the IA-based one, respectively.
	This result demonstrate that for multi-antenna communication systems, the proposed RA-based scheme can achieve significant performance improvement even when the number of RAs is small.

 	\vspace{-0.2cm}
 	\section{Conclusion}
 	\vspace{-0.1cm}
 	
 	This paper investigated the RA-enhanced single/multi-beam forming by leveraging the new spatial DoFs via the antennas' rotation optimization. 
 	This paper aimed to maximize the minimum array gain over signal directions with a given constraint on the maximum array gain over interference directions. 
 	For the special case of single-beam forming, the ARAV was derived in closed-form with MRC beamformer applied to the AWV. 
 	For the general case of multi-beam forming, an AO algorithm was proposed to alternately optimize the ARAV and the AWV in an iterative manner.
 	Simulation results showed that the proposed RA-based scheme can significantly increase the array gain in signal directions while effectively suppressing interference, outperforming both the conventional FOA-based and IA-based schemes.
 	Additionally, for multi-antenna communication systems, the proposed scheme consistently achieves significant performance gain for different numbers of RAs.
 	
 	\vspace{-0.2cm}
 	
	\ifCLASSOPTIONcaptionsoff
		\newpage
	\fi
 	 
	\bibliographystyle{IEEEtran} 	 
	\bibliography{bib_file}

 \end{document}